\documentclass[a4paper,twocolumn]{Gaia2004} 
\usepackage{times}      
\usepackage{epsfig}     
\usepackage{natbib}     
\title{Refining the relativistic model for Gaia: cosmological effects in the BCRS}

\author{Sergei~A.~Klioner,\ }

\author{Michael~H.~Soffel}

\affil{Lohrmann Observatory, Dresden Technical University, Mommsenstr 13, 01062 Dresden, Germany}

\def\aj{AJ}%
\def\prd{Phys.~Rev.~D}%
\def\ve#1{{\bf #1}}

\begin{document}

\setcounter{page}{305}

\keywords{relativity, cosmology, Gaia reference frame}

\maketitle

\begin{abstract}

This paper represents a first attempt of embedding the Barycentric Celestial
Reference System, the fundamental relativistic reference system to be used for
the modeling of
Gaia observations, into some cosmological background. The general
Robertson-Walker metric is transformed into local coordinates where the
cosmological effects are represented as tidal potentials. A version of
a cosmological BCRS is then suggested to lowest order. The effects of
cosmological background on the motion of the solar system are estimated and found
to be completely negligible. The relation to the de Sitter and Schwarzschild-de
Sitter solutions is discussed.

\end{abstract}

{
\vskip -13.7cm

\vbox to 2cm{
\hbox{In: Proceeding of Symposium ``The Three-Dimensional Universe with Gaia'',}
\hbox{\phantom{In:\ }\ 4--7 October 2004, Paris, France (ESA SP-576, January 2005), pp. 305--309}
}

\vskip 12cm

}

\section{Introduction}

The relativistic model for GAIA published by \citet{Klioner:2003} is formulated in
the first post-Newtonian approximation under the assumption that the solar
system is isolated. The relativistic reference system where the positions and
other parameters of the sources are described is the Barycentric Celestial
Reference System (BCRS) of the International Astronomical Union \citep{Soffel:et:al:2003}.
This model can be refined in two principal ways:
%
(1)
one can account for various sources of  gravitational  fields
generated outside the Solar system (cosmology, microlensing, and
gravitational waves);
(2) one can optimize the post-Newtonian formulas to allow a faster numerical
implementation of the model.
%
\noindent
Several projects aimed at such refinements are underway. Here we discuss the
imbedding of the BCRS into some cosmological background.

\section{The BCRS and external gravitational fields}

The BCRS is constructed under the assumption that the solar system is
isolated. One neglects herewith two kinds of effects:
\begin{itemize}
\item[(1)]
tidal forces due to any particular external body (e.g., nearby stars or the
Galaxy),
\item[(2)]
effects of cosmological background.
\end{itemize}
\noindent
The tidal forces can be easily taken into account using the same theoretical
framework as that used to construct the BCRS (e.g. \citet{Soffel:et:al:2003}).
These effects were estimated to be negligible even for GAIA accuracy and,
thus, the corresponding terms have been dropped in the BCRS metric tensor. To
include effects from the cosmological background a new approach will be
necessary.

\section{The Robertson-Walker metric in the vicinity of an observer}
\label{Section-RW-local}

Let us first consider a general Robertson-Walker metric in the neighborhood
of an observer, the size of the neighborhood being about the size of the
solar system. Thus, assuming the usual conditions for
homogeneity and isotropy the cosmological metric can be shown to coincide with the
well-known Robertson-Walker metric which in spherical coordinates
$(x^0=c\,t,\rho,\theta,\varphi)$ reads
\begin{eqnarray}\label{RW-nonflat-nonisotropic}
ds^2&=&-c^2\,dt^2+a^2(t)
\nonumber\\
&&\times
\left({d\rho^2\over 1-k\,\rho^2}+\rho^2\,d\theta^2
+r^2\,\sin^2\theta\,d\varphi^2\right),
\end{eqnarray}
\noindent
with $k=-1$, $0$ or $+1$ being the curvature parameter.
By a simple transformation of the radial coordinate
\begin{equation}
\rho=r\,\left(1+{1\over 4}\,k\,r^2\right)^{-1}
\end{equation}
\noindent
this metric can be transformed into isotropic form
\begin{eqnarray}\label{RW-nonflat}
ds^2&=&g_{\alpha\beta}\,dx^\alpha\,dx^\beta,
\nonumber \\
g_{00}&=&-1,
\nonumber \\
g_{0i}&=&0,
\nonumber \\
g_{ij}&=&a^2(t)\,\left(1+{1\over 4}\,k\,r^2\right)^{-2}\,\delta_{ij},
\end{eqnarray}
\noindent
where $(x^0=c\,t,x^i)$, $r=|x^i|$ are Cartesian coordinates.

It is clear that the following coordinate transformations
\begin{eqnarray}\label{RW-nonflat-local-transformation-T}
T&=&t+\sum_{s=1}^\infty\, C_s(t)\,\left({a(t)\,r\over c}\right)^{2s},
\\
\label{RW-nonflat-local-transformation-R}
X^i&=&a(t)\,x^i\,\left(1+\sum_{s=1}^\infty\, D_s(t)\,\left({a(t)\,r\over c}\right)^{2s}\right)
\end{eqnarray}
\noindent
can be used to transform the metric (\ref{RW-nonflat}) into a form (this form
will be called {\sl local} metric below) where the non-Minkowskian terms appear
only as powers of radial distance from the origin
\begin{eqnarray}\label{RW-nonflat-local}
G_{00}&=&-1+\sum_{s=1}^\infty\, A_s(T)\,\left({R\over c}\right)^{2s},
\nonumber \\
G_{0i}&=&0,
\nonumber \\
G_{ab}&=&\delta_{ab}\,\left(1+\sum_{s=1}^\infty\,B_s(T)\,\left({R\over c}\right)^{2s}\right),
\end{eqnarray}
\noindent
where $X^0=c\,T,X^i$ are the local coordinates, $R=|X^i|$, and $A_s$, $B_s$,
$C_s$ and $D_s$ are unknown functions to be determined from the tensorial
transformation law:
\begin{equation}\label{matching}
g_{\varepsilon\lambda}(t,\ve{x})=
{\partial X^\mu\over \partial x^\varepsilon}\,
{\partial X^\nu\over \partial x^\lambda}\,
G_{\mu\nu}(T,\ve{X}).
\end{equation}
\noindent
The matching procedure allows one to derive the unknown functions in a unique way.
Let us introduce the following notations:
\begin{equation}\label{p-k-notation}
p_s={1\over a(t)}\,{d^s\over dt^s}\,a(t),
\end{equation}
\begin{equation}\label{q-notation}
f={k\,c^2\over a^2(t)}.
\end{equation}
\noindent
Then one has
\begin{eqnarray}
\label{RW-nonflat-local-transformation-coefficient}
A_1&=&p_2,
\nonumber\\
A_2&=&{1\over 4}\,(p_1^4-2\,p_1^2\,p_2-p_2^2+f\,(p_1^2-p_2)),
\nonumber\\
B_1&=&-{1\over 2}\,(p_1^2+f),
\nonumber\\
B_2&=&-{1\over 16\,}\,(p_1^4-4\,p_1^2\,p_2-2\,f\,p_1^2-3\,f^2),
\nonumber\\
C_1&=&{1\over 2}\,p_1,
\nonumber\\
C_2&=&{1\over 8}\,p_1\,(2\,p_2-f),
\nonumber\\
D_1&=&{1\over 4}\,p_1^2,
\nonumber\\
D_2&=&{1\over 16}\,p_1^2\,(2\,p_2-f),
\\
\dots\ .
\nonumber
\end{eqnarray}
\noindent
It is clear already from the dimensionality consideration that the
coefficients $A_s$, $B_s$, $C_s$, $D_s$ are homogeneous polynomials of the
differentiation operator ${d/dt}$ acting on $a(t)$ and $f$ weighted as a
term of degree two. The degrees of the polynomials $A_s$, $B_s$, $C_s$, $D_s$
are $2s$, $2s$, $2s-1$ and $2s$.

Note that $A_s$ and $B_s$ are functions of $T$ in (\ref{RW-nonflat-local}).
The above formulas (\ref{RW-nonflat-local-transformation-coefficient}) for
these quantities should be understood in such a way that $t$ should be
formally replaced by $T$ in the right-hand side of each equation. E.g.
$A_1(T)=\left.{{\ddot a(t)}\over a(t)}\right|_{t\to T}$. A program for {\it
Mathematica} written by the authors allows one to perform the matching
procedure to arbitrary order of $r$.

Thus, we have constructed the general Robertson-Walker metric in local
coordinates where the cosmological effects manifest themselves only as tidal
potentials.

\section{A particular case: the flat de Sitter solution}

In the particular case of a flat de Sitter universe (flat empty universe with
a $\Lambda$ term) one has $a(t)=\exp(c\,\sqrt{\Lambda/3}\,t)$ and $k=0$ in
(\ref{RW-nonflat}), so that the metric reads
\begin{eqnarray}\label{deSitter-global}
g_{00}&=&-1,
\nonumber \\
g_{0i}&=&0,
\nonumber \\
g_{ij}&=&\delta_{ij}\,\exp\left(c\,\sqrt{2\Lambda\over 3}\,t\right).
\end{eqnarray}
\noindent
It is also well known that the flat de Sitter solution can be written in
another form
\begin{eqnarray}
\label{ds2-deSitter-standard}
ds^2&=&-A\,c^2\,dT^2+A^{-1}\,d\rho^2+\rho^2\,(d\theta^2+\sin^2\theta d\varphi^2),
\nonumber\\
A&=&1-{1\over3}\,\Lambda\,\rho^2,
\end{eqnarray}
\noindent
which under the transformation
\begin{eqnarray}\label{R-rho-de-Sitter}
R&=&{2\,\rho\over 1+\sqrt{1-{1\over 3}\,\Lambda\,\rho^2}},
\\
\label{rho-R-de-Sitter}
\rho&=&{R\over 1+{1\over 12}\,\Lambda\,R^2},
\end{eqnarray}
\noindent
can be re-written in isotropic form
\begin{eqnarray}
\label{deS-ds2-iso}
\label{deS-G-iso}
ds^2&=&G_{\alpha\beta}\,dX^\alpha\,dX^\beta,
\nonumber\\
G_{00}&=&-{\left(1-{1\over12}\,\Lambda\,R^2\over 1+{1\over12}\,\Lambda\,R^2\right)}^2,
\nonumber\\
G_{0i}&=&0,
\nonumber\\
G_{ab}&=&\delta_{ab}\,{1\over \left(1+{1\over12}\,\Lambda\,R^2\right)^2}.
\end{eqnarray}
\noindent
It is easy to see that (\ref{deS-G-iso}) coincides with
(\ref{RW-nonflat-local}) in the case
$a(t)=\exp\left(c\,\sqrt{\Lambda/3}\,t\right)$ and $k=0$.

It is also known that the transformation
\citep{Lemaitre:1925,Robertson:1928,Tolman:1934}
\begin{eqnarray}
\label{t-T-de-Sitter}
t&=&T+{1\over 2c}\,\sqrt{3\over \Lambda}\,\log\left(1-{1\over3}\,\Lambda\,\rho^2\right),
\\
\label{r-rho-de-Sitter}
r&=&{\rho\over \sqrt{1-{1\over3}\,\Lambda\,\rho^2}}\,\exp\left(-\sqrt{\Lambda\over 3}\ c\,T\right)
\end{eqnarray}
\noindent
can be used to transform (\ref{ds2-deSitter-standard}) into
(\ref{deSitter-global}) and vice verse. Inverting
(\ref{t-T-de-Sitter})--(\ref{r-rho-de-Sitter}) one gets
\begin{eqnarray}
\label{T-t-de-Sitter}
T&=&t-{1\over 2c}\,\sqrt{3\over \Lambda}\,\log\left(1-{1\over3}\,\Lambda\,a^2\,r^2\right),
\\
\label{rho-r-de-Sitter}
\rho&=&a\,r
\end{eqnarray}
\noindent
with $a(t)=\exp\left(c\,\sqrt{\Lambda/3}\,t\right)$. Combining
(\ref{T-t-de-Sitter})--(\ref{rho-r-de-Sitter}) with (\ref{rho-R-de-Sitter})
one gets the transformation between the two isotropic versions of the flat
de Sitter metric (\ref{deSitter-global}) and (\ref{deS-G-iso}):
\begin{eqnarray}
\label{T-t-de-Sitter-1}
T&=&t-{1\over 2c}\,\sqrt{3\over \Lambda}\,\log\left(1-{1\over3}\,\Lambda\,a^2\,r^2\right),
\\
\label{R-r-de-Sitter}
R&=&{2\,a\,r\over 1+\sqrt{1-{1\over 3}\,\Lambda\,a^2\,r^2}}.
\end{eqnarray}
\noindent
The coefficients of the series expansions of
(\ref{T-t-de-Sitter-1})--(\ref{R-r-de-Sitter}) coincides with the
corresponding coefficients
(\ref{RW-nonflat-local-transformation-coefficient}) in
(\ref{RW-nonflat-local-transformation-T})--(\ref{RW-nonflat-local-transformation-R}).
This gives an additional
check of the general formulas given in Section \ref{Section-RW-local}.

\section{The cosmological BCRS metric to lowest order}

The cosmological BCRS metric can be considered as a perturbation of the
corresponding cosmological metric. The gravitational field of the central
body (or a system of bodies, e.g. of the solar system) should be embedded
into the cosmological metric. Neglecting the interaction of the cosmological
fluid (including the cosmological constant) with the solar system matter one
gets the following simple version of the cosmological BCRS metric:
\begin{eqnarray}\label{BRS:metric}
g_{00}&\approx&-1+{2\over c^2}\,w(t,\ve{x})-{2\over c^4}\,w^2(t,\ve{x})
+{1\over c^2}\,A_1(t)\,|\ve{x}|^2,
\nonumber \\
g_{0i}&\approx&-{4\over c^3}\,w^i(t,\ve{x}),
\nonumber \\
g_{ij}&\approx&\delta_{ij}\left(1+{2\over c^2}\,
w(t,\ve{x})+{1\over c^2}\,B_1(t)\,|\ve{x}|^2\right),
\end{eqnarray}
\noindent
where $w$ and $w^i$ are the usual BCRS metric potentials. Here we neglected:
%
%
(1)
higher post-Newtonian terms ${\cal O}(c^{-5})$ in $g_{00}$ and in $g_{0i}$,
and ${\cal O}(c^{-4})$ in $g_{ij}$ due to post-post-Newtonian (and higher order)
effects from the solar system matter;
%
(2)
higher-order cosmological terms ${\cal O}(|\ve{x}|^4)$ proportional to
$A_s$ and $B_s$ with $s\ge2$,
%
(3)
any terms induced by the interaction between the solar system matter and the
cosmological fluid.
%
%
\noindent
For the case when the only source of energy-momentum tensor is the solar
system matter and the field equations contain the cosmological constant
$\Lambda$ \citet{Soffel:Klioner:2003} have explicitly demonstrated how the metric
(\ref{BRS:metric}) can be derived from the field equations in the
corresponding approximation.

\section{The local expansion hypothesis}

The apparently simple question whether the cosmological expansion
happens also locally (say, whether a hydrogen atom or the Solar system
also expand) is a very complicated one and presents still an unsolved
problem. Starting from Einstein himself different authors got different
answers using different arguments (see \citet{Bonnor:2000} for a review
of recent progress). Certainly, the answer to this question crucially
depends on our model of the matter in the universe.

The cosmological BCRS metric suggested above implies that the
cosmological expansion has certain influence on the properties of space-time
within the solar system. This is certainly true for the terms coming from the
cosmological constant $\Lambda$ or "vacuum energy" since this energy source is
present everywhere. Note that recent cosmological observations suggest that
about 73\% of the energy in the universe comes from that source. The
applicability of the suggested BCRS metric to the 4\% coming from the luminous
matter and the 23\% of the dark matter has to be investigated further. Ideas
of cosmological averaging might be employed here \citep{Futamase:1996}.

\section{The Schwarzschild-de Sitter solution}

The well-known Schwarzschild-de Sitter solution provides additional arguments
in justifying the cosmological BCRS metric. The standard form of the
Schwarzschild-de Sitter metric resembles (\ref{ds2-deSitter-standard}):
\begin{eqnarray}
\label{ds-Schwarzschild-de-Sitter}
ds^2&=&-{\cal A}\,c^2\,dT^2+{\cal A}^{-1}\,d\rho^2+\rho^2(d\theta^2+\sin^2\theta d\varphi^2),
\nonumber \\
{\cal A}&=&1-{2m\over \rho}-{1\over3}\,\Lambda\,\rho^2.
\end{eqnarray}
\noindent
This solution corresponds to the Schwarzschild solution with mass $m$ for the field
equations with cosmological constant. For $\Lambda=0$ this solution coincides
with the Schwarzschild solution in standard coordinates, and for $m=0$ this agrees with
the de Sitter solution (\ref{ds2-deSitter-standard}).
\citet{Robertson:1928} has shown that the metric (\ref{ds-Schwarzschild-de-Sitter})
under the transformations (in this section $a(t)=\exp(c\,\sqrt{\Lambda/3}\,t)$)
\begin{eqnarray}
\label{T-t-Schwarzschild-de-Sitter-closed-isotropic-global}
T&=&t+{8m^2\over c}\,\sqrt{\Lambda\over 3}\ F\left({1-{m\over 2\,a(t)\,r}\over 1+{m\over 2\,a(t)\,r} }\right),
\\
\label{F(x)}
F(x)&=&\int{dx\over (1-x^2)\left(x^2\,(1-x^2)^2-{4\over 3}\,m^2\,\Lambda\right)},
\\
\label{rho-r-Schwarzschild-de-Sitter-closed-isotropic-global}
\rho&=&a(t)\,r\,{\left(1+{m\over 2a(t)\,r}\right)}^2
\end{eqnarray}
\noindent
can be put into form
\begin{eqnarray}
\label{g-Schwarzschild-de-Sitter-closed-isotropic-global}
ds^2&=&g_{\alpha\beta}\,dx^\alpha\,dx^\beta,
\nonumber\\
g_{00}&=&-{\left(1-{m\over 2\,a(t)\,r}\over 1+{m\over 2\,a(t)\,r}\right)}^2,
\nonumber\\
g_{0i}&=&0,
\nonumber\\
g_{ij}&=&\delta_{ij}\,a(t)^2\,{\left(1+{m\over 2\,a(t)\,r}\right)}^4.
\end{eqnarray}
\noindent
For $\Lambda=0$ this metric coincides with the Schwarzschild solution in
isotropic coordinates, and for $m=0$ this metric gives the de Sitter solution
in the form (\ref{deSitter-global}). It is important to note that Eqs.
(\ref{T-t-Schwarzschild-de-Sitter-closed-isotropic-global})--(\ref{rho-r-Schwarzschild-de-Sitter-closed-isotropic-global})
coincide with (\ref{T-t-de-Sitter})--(\ref{rho-r-de-Sitter}) for $m=0$.

Metric (\ref{ds-Schwarzschild-de-Sitter}) written in isotropic or harmonic
gauge represents an exact solution of the cosmological BCRS in local
coordinates (cosmological terms as tidal potentials only) for the particular
case of the de Sitter metric with a spherically symmetric perturbation. This
solution can be used to verify the cosmological BCRS metric in higher
approximations. The details will be published elsewhere.

\section{Dynamical cosmological effects in the Solar system}

The cosmological terms in the suggested BCRS metric lead to an additional
central force described by the disturbing function
\begin{equation}\label{cosm-perturbation}
R={1\over 2}\,A_1\,r^2.
\end{equation}
\noindent
The parameter $A_1$ can be calculated as
\begin{equation}
A_1=p_2={\ddot a\over a}=-q\,H^2,
\end{equation}
\noindent
where $H$ and $q$ are the Hubble constant and the deceleration parameter,
respectively. Using the current best estimates of the parameters
$q\approx-0.6$ and $H\approx71$ km/s/Mpc one gets
\begin{equation}
A_1\approx3.2\times10^{-36}\ {\rm s}^{-2}.
\end{equation}
\noindent
This is the value of $A_1$ at the present moment. The time dependence of this
quantity is different for different cosmological models. In the flat de Sitter universe $A_1$
is time independent: $A_1={1\over 3}\,c^2\,\Lambda={\rm const}$. In general
$A_1$, however, is time dependent. At the distance of Pluto ($r = 40$ AU) the maximal
value of the disturbing cosmological acceleration amounts to $A_1\times 40\
{\rm AU}\ \approx\ 2\times10^{-23}$ m/s$^2$. Let us check how the cosmological
perturbations under study change the orbit of planets. Since the disturbing function
(\ref{cosm-perturbation}) gives a
central force, the osculating inclination $i$, osculating longitude of the
node $\Omega$, and osculating semi-latus rectum $p=a(1-e^2)$,  $a$ and $e$
being the semi-major axis and the eccentricity, remain constant:
\begin{eqnarray}
\label{i-const}
i&=&{\rm const},
\\
\label{Omega-const}
\Omega&=&{\rm const},
\\
\label{p-const}
p&=&a\,(1-e^2)={\rm const}.
\end{eqnarray}
\noindent
Since the time scale over which $A_1$ is changing is much
larger than typical orbital periods in the solar system, we might use an
adiabatic approximation and put $A_1={\rm const}$ for the calculations of the other
osculating elements. In this case one gets that the secular changes of
averaged argument of perihel\-ion $\overline{\omega}$ and averaged mean anomaly
$\overline{M}_0$ read
\begin{eqnarray}
\Delta\overline{\omega}&=&{3\sqrt{1-e^2}\over4\,\pi}\,A_1\,P^2\ \hbox{\rm rad/revolution},
\\
\Delta\overline{M}_0&=&-{7+3e^2\over4\,\pi}\,A_1\,P^2\ \hbox{\rm rad/revolution}.
\end{eqnarray}
\noindent
Here and below $P$ is the period of motion of the body under consideration.
There are no secular effects in the semi-major axis and eccentricity
and the amplitudes of periodic effects in these elements may amount to
\begin{eqnarray}
\Delta_e&=&{1-e^2\over 4\,\pi^2}\,A_1\,P^2,
\\
\Delta_a/a&=&{2e\over 1-e^2}\,\Delta_e={e\over 2\pi^2}\,A_1\,P^2.
\end{eqnarray}
This shows that the secular perturbations in the argument of perihelion
$\omega$ and in the mean anomaly $M$ for Pluto are about $10^{-5}$
microarcsecond per century and three orders of magnitude smaller for Mercury.
The amplitudes of the periodic effects in $a$ and $e$ are about $10^{-17}$
for Pluto and $10^{-23}$ for Mercury. Therefore, these cosmological
perturbations are completely negligible for the solar system dynamics over times
much less than the age of the Universe.

\section{Further work}

Let us mention two almost independent problems which should still be solved:
%
%

(1) A better version of the cosmological BCRS metric should be constructed where
the non-linearity of the field equations is taken into account.
%

(2) Parameters (e.g., parallaxes of distant sources) obtained while processing
observations with the standard (non-cosmological) BCRS should be related to
the parameters  meaningful in cosmological context. Let us note here, that
the parallax distance defined as $d_p\approx a(t)\,r\approx R$ agrees with the
coordinate distance in our local coordinates.
%

\end{document}